\documentclass[aps,preprint]{revtex4}%
\usepackage{amsfonts}
\usepackage{amsmath}
\usepackage{amssymb}
\usepackage{graphicx}%
\newcommand{\be}{\begin{equation}}
\newcommand{\ee}{\end{equation}}

\newcommand{\bea}{\begin{eqnarray}}
\newcommand{\eea}{\end{eqnarray}}
\newcommand{\om}{\omega}
\newcommand{\Om}{\Omega}
\newcommand{\br}{{\bf r}}
\newcommand{\cE}{{\cal E}}
\newcommand{\bkk}{{{\bf k},\lambda}}
\begin{document}
\preprint{ }
\title{Influence of radiative damping on the optical-frequency susceptibility}
\author{P. W. Milonni}
\affiliation{Theoretical Division (T-DOT), Los Alamos National Laboratory, Los
Alamos, New
Mexico \ \ 87545}
\author{Robert W. Boyd}
\affiliation{Institute of Optics, University of Rochester, Rochester, New York \
\ 14627}
\begin{abstract}
Motivated by recent discussions concerning the manner in which damping appears
in  the electric polarizability, we show that (a) there is a dependence of the
nonresonant contribution on the damping and that (b) the damping enters
according to the ``opposite sign prescription." We also
discuss the related question of how the damping rates in the polarizability are
related to energy-level decay rates.

\end{abstract}
\volumeyear{year}
\volumenumber{number}
\issuenumber{number}
\eid{identifier}
\date[Date text]{date}
\received[Received text]{date}

\revised[Revised text]{date}

\accepted[Accepted text]{date}

\published[Published text]{date}

\startpage{1}
\endpage{ }
\maketitle

\section{Introduction}
Several recent papers address the question of how material damping effects
should be
included in the response of an atom or molecule to an applied electric field
\cite{and1}-\cite{andn}.
We will consider the simplest case, that of the linear atomic polarizability,
which in the
absence of damping is given by the Kramers-Heisenberg formula,
\be
\alpha_i(\om)={e^2\over
3\hbar}\sum_j|\br_{ji}|^2\left({1\over\om_{ji}-\om}+{1\over\om_{ji}+\om}\right)
\label{kh1}
\ee
for state $i$. Here $\om_{ji}$ and $\br_{ji}$ are the transition (angular)
frequency and coordinate matrix element,
respectively, between states $i$ and $j$, and the field frequency $\om$ is
assumed to be far removed
from any of the atomic transition frequencies $\om_{ji}$. More generally one
associates damping rates
$\gamma_{ji}$ with the different transitions and writes
\be
\alpha_i(\om)={e^2\over
3\hbar}\sum_j|\br_{ji}|^2\left({1\over\om_{ji}-\om-i\gamma_{ji}}+
{1\over \om_{ji}+\om+i\xi\gamma_{ji}}\right)
\label{kh2}
\ee
where $\xi=+1$ according to the so-called ``opposite sign" prescription and
$\xi=-1$ in the
``constant sign" prescription. The difference appears only in the
nonresonant terms, and is therefore unimportant in most situations. However,
the question of which prescription is the correct one raises some interesting
points, as we shall see, and  
the purpose of this paper is to address some of these points as well as to
answer the question of
whether one should take $\xi=+1$ or $\xi=-1$ in equation (\ref{kh2}).

One might ask first whether a damping term should appear at all in the
nonresonant part
of the Kramers-Heisenberg formula, i.e., whether we should in fact take $\xi=0$
instead of
either $\xi=+1$ or $\xi=-1$. An analysis involving the diagonalization of the
(two-level) atom-field
Hamiltonian in the rotating-wave approximation, for instance, shows that there
is
 no damping term in the nonresonant denominator \cite{loudon}, a result that is
certainly accurate for most practical 
purposes. In a broader context the issue here is an old one. Thus
the imaginary part of the polarizability (\ref{kh2}) implies an absorption
coefficient having the
usual Lorentzian form
\begin{eqnarray}
{\gamma\over (\om-\om_0)^2+\gamma^2} 
\label{rline}
\end{eqnarray}
as well as a nonresonant part 
\begin{eqnarray}
{\gamma\over (\om+\om_0)^2+\gamma^2}
\label{arline}
\end{eqnarray}
for a transition of frequency $\om_0$ and linewidth $\gamma$, and one might
question whether, as a matter
of principle, (\ref{arline}) contributes to the absorption lineshape. In his
consideration of possible
corrections to the Weisskopf-Wigner lineshape, Lamb \cite{lamb} noted that
``such a contribution [as (\ref{arline})]
appears in some derivations," but added that it would be negligible compared
with the resonant
contribution (\ref{rline}).

The effect of damping on the nonresonant part of the polarizability is not an
entirely trivial matter, and
the literature relating to the subject reveals significant disagreement on some
rather basic aspects
of dissipation theory. The purpose of this paper is to address the principal
points where there is disagreement
and to obtain what we regard as the correct form of the polarizability when
damping is included.
    
In the following section we consider the problem of the electric-dipole
interaction of a two-level atom with the
quantized electromagnetic field, assuming that all but one of the field modes
are initially unoccupied. Using
the rotating-wave approximation (RWA) for the atomic source field but not for
the applied  field,
 we obtain exactly the result cited earlier \cite{loudon}, and in particular we
find that there is no contribution from 
(radiative) damping to
the nonresonant term in the polarizability. In Section \ref{sec:nonrwa} we go
beyond the RWA in the atomic source field
and find that the damping now appears in the nonresonant term, and that it does
so in accordance with the
``opposite sign" prescription. Section \ref{sec:disc} presents a discussion of
these results, including their connection 
to the classical theory of radiative damping.
Section \ref{sec:sum} focuses on the form of the damping rate $\gamma_{ji}$, and
we argue that, contrary to what 
sometimes appears in the literature, $\gamma_{ji}$ depends on the sum rather
than the difference of the
level decay rates. Our conclusions are summarized in Section \ref{sec:summary}. 

\section{\label{sec:rwa}Derivation of Linear Polarizability: RWA}
The model we consider is described by the Hamiltonian
\be
H=\hbar\om_0\sigma_z+\sum_k\hbar\om_ka_k^{\dag}a_k-d E\sigma_x 
=\hbar\om_0\sigma_z+\sum_k\hbar\om_ka_k^{\dag}a_k
-i\hbar\sum_kC_k(a_k-a_k^{\dag})(\sigma +\sigma^{\dag})
\label{rwa1}
\ee 
for a two-level atom (TLA) with transition frequency $\om_0$ and dipole moment
$d$ interacting with the
electromagnetic field. The $\sigma$ operators are the usual Pauli operators,
with $\sigma$ and $\sigma^{\dag}$
the lowering and raising operators for the TLA. $a_k$ and $a_k^{\dag}$ are the
annihilation and creation
operators for field mode $k$, and $C_k=({\bf d}\cdot{\bf
e}_{\bkk}/\hbar)(2\pi\hbar\om_k/V)^{1/2}$,
with $V$ the quantization volume. The subscript $k$ denotes $(\bkk)$, where
${\bf k}$ is the wave vector
assocated with a plane-wave mode of frequency $\om_k=|{\bf k}|c$ and ${\bf
e}_{\bkk}$ is a corresponding
polarization unit vector (${\bf k}\cdot{\bf e}_{\bkk}=0, {\bf e}_{\bkk}\cdot{\bf
e}^*_{{\bf k},\lambda'}=
\delta_{\lambda\lambda'},\lambda=1,2$).

 The commutation relations for the atom and field operators give the
 Heisenberg equations of motion
\be
\dot\sigma=-i\om_0\sigma+\sum_kC_k(\sigma_za_k-a_k^{\dag}\sigma_z)
\label{rwa3}
\ee
\be
\dot{a}_k=-i\om_ka_k+C_k(\sigma +\sigma^{\dag})
\label{rwa4}
\ee
We have chosen a normal ordering for the field annihilation and creation
operators, which is especially useful in the case that the applied field is
described by a coherent state [Eq. (\ref{rwa8})].
As we are interested only in determining the linear response, the equation of
motion for $\sigma_z$ will not be needed for our purposes.

 The formal solution of equation (\ref{rwa4}) is
\be
a_k(t)=a_k(0)e^{-i\om_kt}+C_k\int_0^tdt'[\sigma(t')+\sigma^{\dag}(t')]e^{i\om_k(t'-t)}
\label{rwa5}
\ee
 In one version of the rotating-wave approximation (RWA) we ignore the coupling
between the creation operator
for the field and the raising operator for the atom; this corresponds, for the
purpose of obtaining the equation of motion
for the field operators, to the neglect of the terms
$a_k^{\dag}\sigma^{\dag}$ and $\sigma a_k$ in the Hamiltonian (\ref{rwa1}). In
this approximation
the equation of motion for $\sigma$ becomes
\bea
\dot\sigma(t)&=&-i\om_0\sigma(t)
+\sum_kC_k[\sigma_z(t)a_k(0)e^{-i\om_kt}-a_k^{\dag}(0)\sigma_z(t)e^{i\om_kt}]
\nonumber \\
&&\mbox{}+\sum_kC_k^2\int_0^tdt'\sigma_z(t)\sigma(t')e^{i\om_k(t'-t)}
\label{rwa6}
\eea
Note that we are {\it not} making an RWA in the free-field operators $a_k(0)$
and $a_k^{\dag}(0)$, so
that both annihilation and creation free-field operators [$a_k(0)$ and
$a_k^{\dag}(0)$] appear in 
(\ref{rwa6}). 

Taking expectation values over the initial atom-field state on both sides of
(\ref{rwa6}), we have
\bea
\langle\dot\sigma(t)\rangle&=&-i\om_0\langle\sigma(t)\rangle
+\sum_kC_k[\langle\sigma_z(t)a_k(0)\rangle
e^{-i\om_kt}-\langle a_k^{\dag}(0)\sigma_z(t)\rangle e^{i\om_kt}] \nonumber \\
\mbox{}&&+\sum_kC_k^2\int_0^tdt'\langle\sigma_z(t)\sigma(t')\rangle
e^{i\om_k(t'-t)}
\label{rwa7}
\eea
We assume that the initial field state $|\psi_F\rangle$ corresponds to a single
occupied mode described by a coherent state with
\be
a_k(0)|\psi_F\rangle=\alpha|\psi_F\rangle \ , \ \ \ \
\langle\psi_F|a_k^{\dag}(0)=\alpha^*\langle\psi_F|
\label{rwa8}
\ee
corresponding to the expectation value
\be
\langle E(t)\rangle=i\left({2\pi\hbar\om\over V}\right)^{1/2}[\alpha e^{-i\om
t}-\alpha^*e^{i\om t}]\equiv
{\cal E}_0\cos\om t
\label{rwa9}
\ee
of the applied electric field. Thus
\be
\sum_kC_k[\langle\sigma_z(t)a_k(0)\rangle
e^{-i\om_kt}-\langle a_k^{\dag}(0)\sigma_z(t)\rangle e^{i\om_kt}]
=-{i\over\hbar}d_x\cE_0\cos\om t
\langle\sigma_z(t)\rangle
\label{rwa10}
\ee
where $d_x$ is the component of the dipole matrix element along the direction of
the applied field; 
$d_x^2=d^2/3$ for the spherically symmetric atom.

We assume that the operator $\sigma_z$, which corresponds to the population
inversion, changes
sufficiently slowly that we may take
\be
\langle\sigma_z(t)\sigma(t')\rangle\cong\langle\sigma_z(t')\sigma(t')\rangle=-\langle\sigma(t')\rangle
\label{rwa13}
\ee
in the integral appearing in (\ref{rwa7}). Since we want to obtain the
polarizability for the TLA in a particular 
state, we assume further that the atom remains with high probability in its
initial state. Assuming this
initial state to be the lower state, we approximate $\langle\sigma_z(t)\rangle$
by $-1$, so that,
using (\ref{rwa10}) and the approximation (\ref{rwa13}), we replace (\ref{rwa7})
by
\be
\langle\dot\sigma(t)\rangle=-i\om_0\langle\sigma(t)\rangle
+i{d\over\hbar}d_x\cE_0\cos\om t-
\sum_kC_k^2\int_0^tdt'\langle\sigma(t')\rangle e^{i\om_k(t'-t)}
\label{rwa11}
\ee

We seek a solution of (\ref{rwa11}) of the form
\be
\langle\sigma(t)\rangle=se^{-i\om t}+re^{i\om t}
\label{rwa12}
\ee
with $s$ and $r$ constants to be determined. This implies
\bea
i(\om_0-\om)se^{-i\om t}+i(\om_0+\om)re^{i\om t}&=&{id_x\over
2\hbar}\cE_0(e^{-i\om t}
+e^{i\om t})-[\gamma_-(\om)-i\Delta_-(\om)]se^{-i\om t} \nonumber \\
&&\mbox{}-[\gamma_+(\om)-i\Delta_+(\om)]re^{i\om t}
\eea
where 
\bea
\gamma_{\pm}(\om)&=&{\cal{R}}e\sum_kC_k^2\int_0^tdt'e^{i(\om_k\pm\om)(t'-t)}\rightarrow
{V\over 8\pi^3}{2\pi\over\hbar V}\int d^3k\om_k\sum_{\lambda}|{\bf d}\cdot{\bf
e}_{\bkk}|^2\pi\delta(\om_k\pm\om)
\nonumber \\
&=&{d^2\over 4\pi^2\hbar c^3}\int
d\Om\Om^3\pi\delta(\Om\pm\om)\int_0^{\pi}d\theta\sin^3\theta \nonumber \\
&=&{2d^2\over 3\hbar
c^3}\int_0^{\infty}d\Om\Om^3\delta(\Om\pm\om)={2d^2\om^3\over 3\hbar
c^3}U(\pm\om)
\label{rwa15}
\eea
and
\be
\Delta_{\pm}(\om)={2d^2\over 3\pi\hbar c^3}{\rm
P}\int_0^{\infty}{d\Om\Om^3\over\Om\pm\om}
\label{rwa16}
\ee
for $t>>1/\omega$, where $U$ is the unit step function. Note that the
damping rate $\gamma_-(\om)$ is frequency-dependent \cite{remark2}. 
($\Delta_{\pm}(\om)$ is obviously divergent but, as discussed in Section 4,
this
has no direct bearing on our conclusions regarding the effect of damping on the
polarizability.)
To obtain the polarizability $\alpha(\om)$ we write
\be
p=d_x\langle\sigma_x\rangle=d_x(\langle\sigma\rangle +
\langle\sigma^{\dag}\rangle) 
=2d_x{{\cal R}}{\rm e}[(r+s^*)e^{-i\om t}] 
\equiv {{\cal R}}{\rm e}[\alpha(\om)\cE_0e^{-i\om t}]
\label{rwa17}
\ee
for the induced dipole moment. This yields 
\be
\alpha(\om)={d^2\over 3\hbar}\left({1\over
\om_0-\om-\Delta_-(\om)-i\gamma_-(\om)}+{1\over
\om_0+\om-\Delta_+(\om)+i\gamma_+(\om)}\right)
\label{rwa18}
\ee
 Note that $\gamma_+(\om)=0$, and that therefore there is no damping
contribution to the 
second (nonresonant) term. $\gamma_-(\om_0)$ is half the radiative decay rate of
the
upper state in the absence of any applied field.

\section{\label{sec:nonrwa}Derivation of Linear Polarizability without RWA}
Let us now recalculate the polarizability, this time retaining both terms inside
the integral of
equation (\ref{rwa5}), i.e., without making the RWA in the (source)
field produced by the atom under consideration. Then (\ref{rwa6}) is replaced by 
\bea
\langle\dot\sigma(t)\rangle&=&-i\om_0\langle\sigma(t)\rangle -i{d_x\over\hbar}
\cE_0\cos\om t\langle\sigma_z(t)\rangle
\nonumber \\
&&\mbox{}+\sum_kC_k^2\int_0^tdt'[\langle\sigma_z(t)\sigma(t')\rangle +
\langle\sigma_z(t)\sigma^{\dag}(t')\rangle]
e^{i\om_k(t'-t)} \nonumber \\
&&\mbox{}-\sum_kC_k^2\int_0^tdt'[\langle\sigma^{\dag}(t')\sigma_z(t)\rangle +
\langle\sigma(t')\sigma_z(t)\rangle]e^{-i\om_k(t'-t)}
\label{nrw1}
\eea
when we take expectation values as before. The approximations tantamount to
(\ref{rwa13}) are
\bea
\langle\sigma_z(t)\sigma(t')\rangle&\cong&\langle\sigma_z(t')\sigma(t')\rangle=-\langle\sigma(t')\rangle
\nonumber \\
\langle\sigma_z(t)\sigma^{\dag}(t')\rangle&\cong&\langle\sigma_z(t')\sigma^{\dag}(t')\rangle=
\langle\sigma^{\dag}(t')\rangle
\nonumber \\
\langle\sigma^{\dag}(t')\sigma_z(t)\rangle&\cong&\langle\sigma^{\dag}(t')\sigma_z(t')\rangle=
-\langle\sigma^{\dag}(t')\rangle
\nonumber \\
\langle\sigma(t')\sigma_z(t)\rangle&\cong&\langle\sigma(t')\sigma_z(t')\rangle=\langle\sigma(t')\rangle
\label{nrw2}
\eea
where we use the equal-time identities
$\sigma_z(t)\sigma(t)=-\sigma(t)\sigma_z(t)=-\sigma(t)$.
Using these approximations in (\ref{nrw1}), together with the approximation
$\langle\sigma_z(t)\rangle\cong -1$
in the second term, we obtain the non-RWA extension of (\ref{rwa11}):
\bea
\langle\dot\sigma(t)\rangle&=&-i\om_0\langle\sigma(t)\rangle+{id_x\over\hbar}\cE_0\cos\om
t
+\sum_kC_k^2\int_0^tdt'[-\langle\sigma(t')\rangle
+\langle\sigma^{\dag}(t')\rangle]e^{i\om_k(t'-t)}
\nonumber \\
&&\mbox{}-\sum_kC_k^2\int_0^tdt'[-\langle\sigma^{\dag}(t')\rangle+\langle\sigma(t')\rangle]e^{-i\om_k(t'-t)}
\label{nrw3}
\eea
It is important to note that in equations (\ref{nrw2}) we have used the
commutation relations between $\sigma_z(t)$
and $\sigma(t)$, $\sigma^{\dag}(t)$, and have obviously not made the
approximation that $\sigma_z$ could be replaced
by $-1$. The latter approximation is made only in the second term of 
(\ref{nrw1}), where $\sigma_z$ multiplies the
applied field but no atom operator, so that the approximation does not violate
the commutation relations from which we
obtained the equations of motion. The two approximations are different: that
made in (\ref{nrw2}) assumes that $\sigma_z(t)$ varies
little on time scales $\sim 1/\om_k$ for field frequencies $\om_k\sim \om$ that
will contribute significantly to the variation of $\langle\sigma(t)\rangle$,
 whereas that made in replacing $\langle\sigma_z(t)\cE_0\cos\om t$ by
$-\cE_0\cos\om t$ assumes that
the atom remains with high probability in its lower state because the field
frequency lies outside the absorption linewidth.
 The difference between these two approximations involving $\sigma_z$ turns out
to be irrelevant for the final results when the RWA is made, 
as is clear from (\ref{rwa13}).

We again have a solution of the form (\ref{rwa12}), now with $s$ and $r$
satisfying
\bea
Xs+Ur^*&=&{d_x\over 2\hbar}\cE_0 \nonumber \\
Vs+Yr^*&=&{d_x\over 2\hbar}\cE_0
\label{nrw33}
\eea
where
\bea
X&=&\om_0-\om-[\Delta_-(\om)-\Delta_+(\om)]-i[\gamma_-(\om)+\gamma_+(\om)]
\nonumber \\
U&=&[\Delta_-(\om)-\Delta_+(\om)]+i[\gamma_-(\om)+\gamma_+(\om)] \nonumber \\ 
Y&=&\om_0+\om+[\Delta_-(\om)-\Delta_+(\om)]+i[\gamma_-(\om)+\gamma_+(\om)]
\nonumber \\
V&=&[\Delta_-(\om)-\Delta_+(\om)]+i[\gamma_-(\om)+\gamma_+(\om)]
\label{nrw4}
\eea
Assuming that $\gamma_{\pm}(\om)$ and $\Delta_{\pm}(\om)$ are small in magnitude
compared to 
$\om_0\pm\om$, we have
\bea
s&\cong&{d_x\cE_0\over 2\hbar}{1\over X}={d_x\cE_0\over 2\hbar}{1\over
\om_0-\om-[\Delta_-(\om)-\Delta_+
(\om)]-i[\gamma_-(\om)+\gamma_+(\om)]} \nonumber \\
r^*&\cong&{d_x\cE_0\over 2\hbar}{1\over Y}={d_x\cE_0\over
2\hbar}{1\over\om_0+\om+[\Delta_-(\om)-\Delta_+
(\om)]+i[\gamma_-(\om)+\gamma_+(\om)]}
\label{nrw5}
\eea
and, from (\ref{rwa17}),
\bea
\alpha(\om)&=&{d^2/ 3\hbar \over \om_0-\om-[\Delta_-(\om)-\Delta_+(\om)]
-i[\gamma_-(\om)+\gamma_+(\om)]} \nonumber \\
&&\mbox{}+{d^2/ 3\hbar\over\om_0+\om+[\Delta_-(\om)-\Delta_+
(\om)]+i[\gamma_-(\om)+\gamma_+(\om)]}
\label{nrw6}
\eea
\section{\label{sec:disc}Discussion}
In contrast to the RWA result (\ref{rwa18}), $\Delta_{\pm}(\om)$ and
$\gamma_{\pm}(\om)$ appear
in both the resonant and nonresonant terms of (\ref{nrw6}). Consider first the
physical significance
of $\Delta_{\pm}(\om)$, assuming that the frequency $\om$ of the initially
occupied field mode is
sufficiently close to $\om_0$ that we may take
\be
\Delta_-(\om)\approx \Delta_-(\om_0)={2d^2\over 3\pi\hbar c^3}{\rm
P}\int_0^{\infty}{d\Om\Om^3\over\Om-\om_0}
\label{d1}
\ee
and focusing only on the resonant term in $\alpha(\om)$. In a more complete
analysis involving the transformation 
from the fundamental minimal coupling form of the Hamiltonian to the electric
dipole form, it is found that 
the additional term $2\pi\int d^3{\bf P}^{\perp}\cdot{\bf P} ^{\perp}$
appearing in the transformed Hamiltonian has the effect of replacing (\ref{d1})
by \cite{pwm1}
\be
\Delta_-(\om_0)\approx{2d^2\om_0^2\over 3\pi\hbar c^3}{\rm
P}\int_0^{\infty}{d\Om\Om\over
\om-\om_0}
\label{d2}
\ee
With this modification it is seen that
$\Delta(\om_0)\equiv\Delta_-(\om_0)-\Delta_+(\om_0)$ is simply the
(unrenormalized) TLA 
radiative frequency shift, i.e., the difference in the radiative {\it level}
shifts of the two levels \cite{vac}. In general, however,
the approximation (\ref{d1}) is not applicable, and the radiative level shifts
$\hbar\Delta_{\pm}(\om)$ depend on
the frequency of the initially occupied mode. In the polarizability (\ref{nrw6})
the frequency shift 
$\Delta(\om)\equiv\Delta_-(\om)-\Delta_+(\om)$ adds to the field frequency $\om$
in both the resonant and nonresonant
terms, whereas in the RWA $\Delta_+(\om)$ does not appear in the resonant term
and $\Delta_-(\om)$
does not appear in the nonresonant term. In other words, the RWA does not
correctly include the radiative
frequency shift as the difference in the radiative level shifts of the TLA.

The expressions for the level shifts $\hbar\Delta_{\pm}(\om)$ are specific to
the TLA model, but are
easily generalized to the case of a real atom. This extension, even with the
standard renormalization
procedures, still leaves us with divergent level shifts in the nonrelativistic
approximation. A high-frequency
cutoff $mc^2/\hbar$ results in Bethe's approximation to the Lamb shift
\cite{vac}. Since
this procedure is very well known, and we are in any case only concerned with
the form in which the
radiative corrections appear in the polarizability, and not their numerical
values, we will simply
assume henceforth that the frequency shift has been accounted for in writing
$\om_0\pm\om$.

Thus
\be
\alpha(\om)={d^2\over 3\hbar}\left({1\over\om_0-\om-i\gamma(\om)}
+{1\over\om_0+\om+i\gamma(\om)}\right)
\label{d3}
\ee
where $\gamma(\om)=\gamma_-(\om)+\gamma_+(\om)$. Like $\Delta(\om)$, 
$i\gamma(\om)$ is effectively an
addition to the applied field frequency $\om$. Unlike the frequency shift,
however, the damping rate 
$\gamma(\om)$ is half the {\it sum} of the decay rates $\gamma_{\pm}(\om)$ of
the two levels. Of course
the decay rate $\gamma_+(\om)$ of the ground state in our two-level model is
zero but, as discussed
in the next section, (\ref{d3}) is valid more generally when the decay rate of
the lower level
of the transition is not zero. That is, the damping rate appearing in the
contribution to the polarizability 
from any given transition involves half the sum of the decay rates of the two
levels of the transition. 

Regardless of whether the lower-level decay rate vanishes, the non-RWA result
(\ref{d3}) shows that
both the resonant and nonresonant contributions to the polarizability have a
nonvanishing damping
term in their denominators, this damping term being half the upper-level decay
rate. In particular, it is seen 
that the damping appears according to the
``opposite sign prescription," i.e., {\it $\xi=+1$ is the correct choice in the
dispersion formula (\ref{kh2})}.
The same conclusion was reached by different lines of reasoning by Buckingham
and Fischer \cite{buck1}.

Note that, if $\gamma$ is taken to be a (positive) constant, independent of
frequency, then the opposite
sign prescription is consistent with the causality requirement that the
polarizability should be analytic
in the upper half of the complex $\om$ plane \cite{nuss}. But in general the
decay rates are in fact frequency-dependent 
\cite{remark2}, and causality is ensured only if the model used to calculate
$\gamma(\om)$ is itself causal.
In fact, as recalled below, radiative damping provides an example in which this
is not the case.

In one approach to a classical calculation of the natural lineshape, one
considers the solution $x(t)=A_0e^{-\gamma t}
\sin(\om_0t)$ of a damped dipole oscillator with resonance frequency $\om_0$.
The lineshape is taken to be proportional to the squared modulus of the Fourier
transform
\be
a(\om)={A_0\over 2\pi}\int_0^{\infty}dte^{-\gamma t}e^{i\om
t}\sin(\om_0t)\propto\left({1\over
\om_0-\om-i\gamma}+{1\over\om_0+\om+i\gamma}\right)
\label{d4}
\ee
and is seen to be consistent with the ``opposite sign" prescription. In contrast
to this, an old
paper by Weisskopf \cite{weissk1} implies the result
\be
a(\om)\propto\left({1\over\om_0-\om-i\gamma}-{1\over\om_0+\om-i\gamma}\right)
\label{d5}
\ee  
which is consistent with the ``constant sign" prescription. However, since this
result is based on the integral appearing in (\ref{d4}),
it seems that (\ref{d5}) involves a sign error or perhaps just a typographical
error.

Since the absorption coefficient may for our purposes be taken to be
proportional to the imaginary part of  $\alpha(\om)$,
equation (\ref{d3}) implies an absorption lineshape proportional to
\be
L(\om)={\gamma\over(\om_0-\om)^2+\gamma^2}-{\gamma\over(\om_0+\om)^2+\gamma^2}
\label{d6}
\ee
The same result, for $\gamma$ taken to be a constant, was obtained on the basis
of the Lorentz model by Van Vleck
and Weisskopf \cite{vanw}, who noted that the minus sign in the nonresonant term
``must be used because the
excitation of the molecule is here accompanied by emission rather than
absorption of a light quantum," a process
which  is excluded when the RWA is made \cite{remark3}.

It is also of interest to compare the result  (\ref{d3}) with the corresponding
result given by the classical
theory of radiative damping based on the equation
\be
\ddot{x}+\om_0^2x-{2e^2\over 3mc^3}\stackrel{...}{x}=e\cE_0\cos\om t
\label{d7}
\ee
The polarizability of the classical dipole oscillator described by this equation
is
\bea
\alpha_{cl}(\om)&=&{e^2/m\over\om_0^2-\om^2-{2\over 3}(ie^2/mc^3)\om^3}
\nonumber \\
&=&{e^2\over
2m\om_0'}\left({1\over\om_0'-\om-i\gamma_{cl}(\om)}+{1\over\om_0'+\om
+i\gamma_{cl}(\om)}\right)
\label{d8}
\eea
where $\om_0'=\sqrt{\om_0^2-\gamma_{cl}^2(\om)}$ and
$\gamma_{cl}(\om)=(e^2/3mc^3)\om^2$.
The replacements $e^2/2m\om_0'\rightarrow e^2f_1/2m\om_0'$ and
$e^2\om^2/3mc^3\rightarrow e^2f_2\om^2/3mc^3$,
where $f_1=2m\om_0'd^2/e^2\hbar$ and $f_2(\om)=(2m\om d^2/e^2\hbar)$, make
the classical result (\ref{d7}) equivalent to (\ref{d3}). These replacements
involving effective oscillator
strengths $f_1$ and $f_2$ are the usual substitutions required to put classical
oscillator results in agreement
with some of the corresponding quantum-mechanical expressions. 

The $\om^3$ in the denominator of (\ref{d8}), or in other words the third
derivative of $x$ in equation (\ref{d7}), leads to
a pole in the upper half of the complex $\om$ plane, thus violating the
causality requirement that the polarizability be
 analytic in the upper half-plane. The nonrelativistic theory of radiative
reaction
 is well known to be acausal, but the acausality occurs on such a short time
scale that relativistic
quantum effects must be taken into account. For most practical purposes the
acausality is of no
consequence. Thus, for instance, equation (\ref{d8}) leads to the correct
extinction coefficient ($\propto \om
{{\cal I}}{\rm m}[\alpha(\om)]$) due to Rayleigh scattering.

\section{\label{sec:sum}Relation of Damping in the Polarizability to Level Decay
Rates}
These considerations are easily extended beyond the two-level model, with the
result that the linear
atomic polarizability has the form
\be
\alpha_i(\om)={1\over
3\hbar}\sum_j|\br_{ji}|^2\left({1\over\om_{ji}-\om-i\gamma_{ji}(\om)}+
{1\over \om_{ji}+\om+i\gamma_{ji}(\om)}\right)
\label{da0}
\ee
The damping rate $\gamma_{ji}(\om)$ has a ``dephasing" contribution associated,
for instance, with elastic
collisions, as well as a contribution associated with the decay rate of the
atomic states $i$ and $j$. Here we consider
only the latter contribution, which is due to radiative decay and other loss
processes. In the case of radiative
decay, for instance, $\gamma_{ji}$ is found, by a straightforward multilevel
generalization of the calculations in
the preceding sections, to be half the sum of the radiative decay rates
associated with the two states $i$ and $j$:
\be
\gamma_{ji}(\om)={2e^2\om^3\over 3\hbar c^3}\left(\sum_{E_j>E_m}|{\bf
r}_{jm}|^2+\sum_{E_i>E_m}
|{\bf r}_{im}|^2\right)
\label{da1}
\ee
where $E_m$ denotes the energy of state $m$. If we replace $\om^3$ by $\om_0^3$
in this formula, we obtain
half the {\it spontaneous} decay rate of state $j$ in the case that the field is
initially
in the vacuum state. This result was obtained, for example, by
Weisskopf and Wigner \cite{ww}, Landau \cite{landau}, and many others
\cite{pwm1}. The same conclusion is reached 
in the more general case where the energy levels decay by nonradiative channels:
$\gamma_{ji}(\om)$ is half the sum of 
the total decay rates of the states $i$ and $j$. 

Various authors, however, have calculated or assumed---erroneously, in our \\
opinion---that $\gamma_{ji}$ involves the 
{\it difference} in the decay rates of the states $i$ and $j$ \cite{sus},
\cite{barron}, \cite{andrews}. In addition to the
Heisenberg-picture calculation leading to the conclusion that $\gamma_{ji}$
involves the sum
rather than the difference of energy-level decay rates, as presented in this
paper, the following simple argument can be used. Let $c_i(t)$ and
$c_j(t)$ be the (Schr\"odinger-picture) probability amplitudes for states $i$
and $j$, and let $\gamma_i$ and
$\gamma_j$ be the decay rates of these states. Then $c_i^*(t)c_j(t)$ and
$c_i(t)c_j^*(t)$, which determine the
polarizability, decay at $\exp[-{1\over 2}(\gamma_i+\gamma_j)t]$, and so the
linewidth in the polarizability must involve
the sum of $\gamma_i$ and $\gamma_j$ rather than the difference. Sushchinskii
\cite{sus}, for instance, expresses
his results in terms of complex energies $E_i'=E_i-{1\over 2}i\Gamma_i$ and
their differences $E_i'-E_j'=
E_i-E_j-{1\over 2}i(\Gamma_i-\Gamma_j)$, whereas the appropriate differences
entering into the polarizability are
$E_i'^*-E_j'$ and $E_i'-E_j'^*$.

Finally we note that Andrews {\it et al.} \cite{andrews} have stated a
polarizability sum rule which in the simplest case
of the linear polarizability can be expressed as $\sum_i\alpha_i(\om)=0$. A
physical plausibility argument for this
sum rule can be adduced as follows. If $p_i$ is the probability that the atom is
in state $i$, then the linear
polarizability at field frequency $\om$ is
\be
\alpha(\om)=\sum_ip_i\alpha_i(\om)
\label{da2}
\ee
Consider the idealized limit in which all the $p_i$ are equal. Then the
polarizability and therefore the induced emission 
or absorption rate at frequency $\om$ becomes proportional to just
$\sum_i\alpha_i(\om)$. But if all the states are equally populated
the net induced emission and absorption rate must vanish, implying the
polarizability sum rule conjectured by Andrews {\it et al.} From the
expression (\ref{da0}) it follows that this sum rule is statisfied only if
$\gamma_{ji}$ is symmetric in $i$ and $j$, i.e.,
$\gamma_{ji}$ must involve the sum rather than the difference of $\gamma_i$ and
$\gamma_j$. (We note that, in the
case of the {\it constant} sign prescription for the damping terms in the
polarizability, the polarizability sum rule would
be satisfied only if $\gamma_{ji}$ were {\it antisymmetric} in $i$ and $j$.) 
 
\section{\label{sec:summary}Summary}
Following a standard, nonrelativisitic approach, we have considered specifically
the case of a two-level atom interacting with the quantized
electromagnetic field, one mode of which is initially occupied and described by
a coherent state.
Working in the Heisenberg picture, we calculated the polarizability with and
without making the
RWA for the atomic source field. In the RWA we obtained a known result, and in
particular
 the nonresonant contribution to the
polarizability was found to have no damping factor in its denominator. Going
beyond the
RWA, however, we found that both the resonant and nonresonant contributions to
the 
polarizability have the radiative damping rate in their denominators, and that
the
polarizability has a form that is consistent with the so-called opposite sign
prescription
for including the damping.

The radiative frequency shift
appearing in the non-RWA expression for the polarizability depends on the
radiative
{\it level} shifts in the correct way, i.e., it is the difference of the two
level
shifts. The damping rate appearing in the non-RWA expression for the
polarizability is half the
{\it sum} of the radiative decay rates of the two levels, in contrast to the
difference of the
decay rates that has been obtained or assumed in some treatments. The fact that
the
polarizability depends symmetrically on the decay rates of the energy levels is 
consistent with the polarizability sum rule of Andrews {\it et al.}
\cite{andrews} when the (correct) opposite
sign prescription is used.
\\ \\
\centerline{\bf Acknowledgement} 
\\ \\
We thank D. L. Andrews, L. C. D\'avila Romero, and G. E. Stedman for helpful
correspondence,
and P. R. Berman and J. H. Carter for useful discussions and suggestions.
R. W. Boyd gratefully acknowledges support by ONR under award N00014-02-1-0797,
by DoE under award DE-FG02-01ER15156,
and by ARO under award DAAD19-01-1-0623.


\begin{thebibliography}{99}
\bibitem{and1} D. L. Andrews, S. Naguleswaran, and G. E. Stedman, Phys. Rev.
A{\bf 57}, 4925 (1998).
\bibitem{buck1} A. D. Buckingham and P. Fischer, Phys. Rev. A{\bf 61}, 035801
(2000).
\bibitem{sted} G. E. Stedman, S. Naguleswaran, D. L. Andrews, and L. C. D\'avila
Romero, Phys. Rev.
A{\bf 63}, 047801 (2001).
\bibitem{buck2} A. D. Buckingham and P. Fischer, Phys. Rev. A{\bf 63}, 047802
(2001).
\bibitem{andn} D. L. Andrews, L. C. D\'avila Romero, and G. E. Stedman, Phys.
Rev. A{\bf 67}, 055801 (2003).
\bibitem{loudon} R. Loudon, {\it The Quantum Theory of Light} (Clarendon Press,
Oxford, 1973),
p. 192. See also G. S. Agarwal and R. W. Boyd, Phys. Rev. A{\bf 67}, 043821
(2003), for a different approach to the
same result.
\bibitem{lamb} W. E. Lamb, Jr., Phys. Rev. {\bf 85}, 259 (1952).
\bibitem{remark2} See G. S. Agarwal and R. W. Boyd, Reference \cite{loudon}.
\bibitem{pwm1} See, for instance, P. W. Milonni, Phys. Rep. {\bf 25}, 1 (1976).
\bibitem{vac} See, for instance, P. W. Milonni, {\it The Quantum Vacuum. An
Introduction to Quantum
Electrodynamics} (Academic Press, San Diego, 1994), Section 4.9.
\bibitem{nuss} See, for instance, H. M. Nussenzveig, {\it Causality and
Dispersion Relations}
 (Academic Press, New York, 1972).
\bibitem{weissk1} V. Weisskopf, Phys. Z. {\bf 34}, 1 (1933).
\bibitem{vanw} J. H. Van Vleck and V. F. Weisskopf, Rev. Mod. Phys. {\bf 17},
227 (1945).
\bibitem{remark3} Van Vleck and Weisskopf went on to show that, when the dipole
orientations or phases
after each collision are treated statistically according to the Boltzmann
distribution, rather than
assumed to be random as in the original Lorentz treatment, one obtains an
absorption lineshape in which the
nonresonant contribution is added rather than subtracted from the resonant term.
Then, instead of the
vanishing absorption predicted by the original Lorentz treatment when the
absorption frequency $\om_0\rightarrow
0$, one obtains the Debye lineshape. 
\bibitem{ww}V. F. Weisskopf and E. Wigner, Z. Phys. {\bf 63}, 54 (1930).
\bibitem{landau}L. D. Landau, Z. Phys. {\bf 45}, 430 (1927).
\bibitem{sus} M. M. Sushchinskii, {\it Raman Spectra of Molecules and Crystals}
(Israel Program for
Scientific Translations, New York, 1972), p. 38.
\bibitem{barron} L. Hecht and L. D. Barron, Mol. Phys. {\bf 79}, 887 (1993);
Chem. Phys. Lett. {\bf 225},
519 (1994).
\bibitem{andrews} D. L. Andrews, L. C. D\'avila Romero, and G. E. Stedman, Phys.
Rev. A{\bf 67}, 055801 (2003).



   
\end{thebibliography}
\end{document}